\documentclass[%
 aps,
 pra,
 superscriptaddress,
 sd,%
 amsmath,amssymb,
 reprint,%
showpacs
]{revtex4-1}
\usepackage{color}
\usepackage{graphicx}% Include figure files
\usepackage{dcolumn}% Align table columns on decimal point
\usepackage{bm}% bold math
\usepackage{soul}
\usepackage{multirow} % Required for multirows
\usepackage{verbatim}

\begin{document}

\title{Germanium microparticles as optically induced oscillators in optical tweezers}

\author{W. H. Campos}
\affiliation{Departamento de F\'isica, Universidade Federal de Vi\c{c}osa, 36570-900 Vi\c{c}osa, MG, Brazil}
\affiliation{Institut f\" ur Physik, Johannes Gutenberg Universit\" at Mainz, 55128 Mainz, Germany}
\author{T. A. Moura}
\affiliation{Departamento de
F\'isica, Universidade Federal de Vi\c{c}osa, 36570-900 Vi\c{c}osa, MG, Brazil}%
\author{O. J. B. J. Marques}
\affiliation{Programa de P\'os-Gradua\c{c}\~ao em Ci\^encia de Materiais, Universidade Federal de Pernambuco, 50740-540 Recife, PE, Brazil}
\author{J. M. Fonseca}\affiliation{Departamento de F\'isica, Universidade Federal de Vi\c{c}osa, 36570-900 Vi\c{c}osa, MG, Brazil}
\author{W. A. Moura-Melo}
\author{M. S. Rocha}
\author{J. B. S. Mendes}
\email{Corresponding author: joaquim.mendes@ufv.br}
\affiliation{Departamento de
F\'isica, Universidade Federal de Vi\c{c}osa, 36570-900 Vi\c{c}osa, MG, Brazil}%
%

%\pacs{Insert PACS}

\date{\today}% It is always \today, today,

\begin{abstract}

\end{abstract}

%\pacs{87.80.Cc, 78.20.-e, 83.85.Ei, 03.65.Vf}% PACS, the Physics and Astronomy
%Use showkeys class option if keyword
\maketitle
%%%%%%%%

\textbf{Oscillatory dynamics is a key tool in optical tweezers applications. It is usually implemented by mechanical interventions that cannot be optically controlled. In this work we show that Germanium semiconductor beads behave as optically induced oscillators when subjected to a highly focused laser beam. Such unusual motion is due to the competition between the usual optical forces \citep{Ashkin111,Ashkin2220,Grier} and the radiometric force related to thermal effects, which pushes the beads from the focal region \cite{Ke}. We characterize the behavior of the Germanium beads in detail and propose a model accounting for the related forces, in good agreement with the experimental data. The well defined direction of oscillations can be manipulated by the polarization of the light beam. Such kind of system can potentially revolutionize the field of optical manipulation, contributing to the design of single molecule machines and the application of oscillatory forces in macromolecules and other soft matter systems.} 

\begin{figure*}
\includegraphics[width=.4\linewidth]{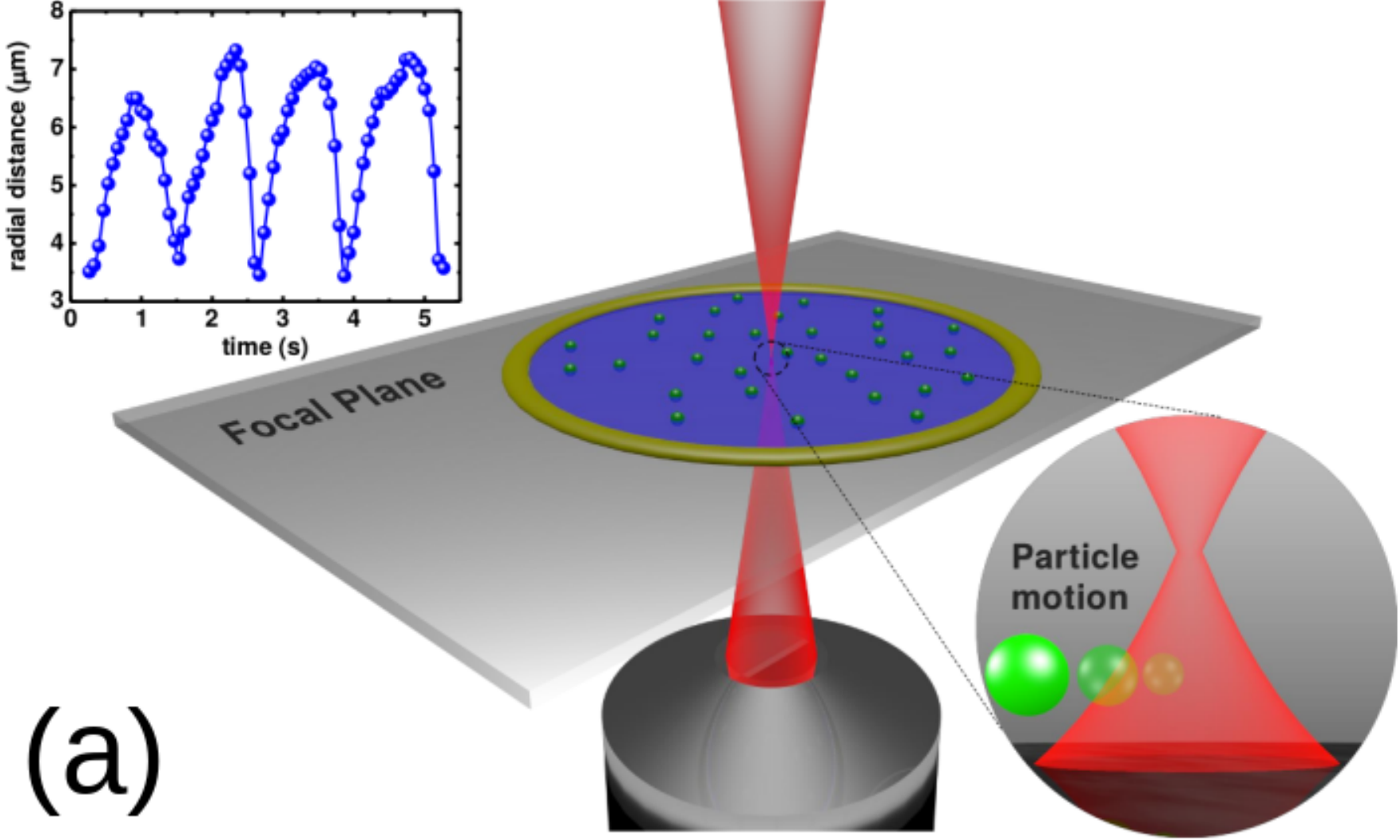}\includegraphics[width=.6\linewidth]{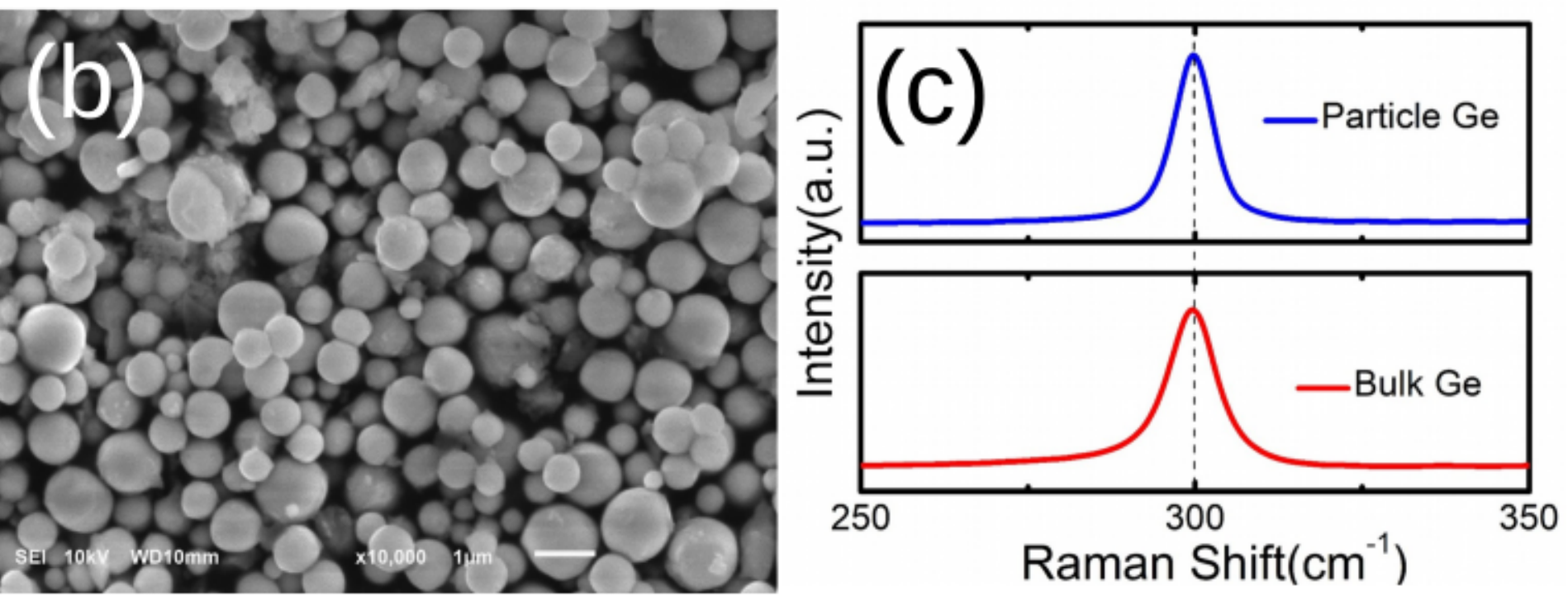}
\caption{(Color online) (a) Experimental setup yielding the optically induced oscillations of Ge microspheres. The beads are suspended in deionized water and subjected to a Gaussian laser beam OT. The sample chamber consists of an {\em o-ring} glued in a microscope coverslip. We have observed that, after placing the beads in the sample chamber with deionized water, they remain suspended around $4\, \mu$m above the coverslip surface, before and during the laser incidence. The inset shows the radial displacement of a bead, relative to the optical axis of the laser. (b) Scanning electron microscopy (SEM) of Ge microspheres obtained by the laser ablation technique. The particles present a well-defined spherical shape with smooth and homogeneous surface. (c) Representative Raman spectra acquired from the Ge-bulk (red color) and a Ge-microparticle (blue color), highlighting the fundamental unstrained Ge Raman line at $ \sim$ 300 $cm^{-1}$, which corresponds to the bulk Ge phonon mode. The Raman spectra indicates that the particles did not experience chemical changes during the synthesis process.}
\label{Fig1}
\label{Fig2}
\end{figure*}

Optical tweezers (OT) works by shedding a highly focused laser beam onto small beads. Radiation pressure and radiometric forces pushes the particle away from the optical axis, while gradient forces, related to refraction effects, attract the object to the laser focus \cite{Ashkin111,Ashkin3330,Ashkin222,taylor2015enhanced}. A metallic particle is usually pushed away, once light is mostly reflected and/or absorbed by the material. On the other hand, a dielectric microparticle suspended in water is usually observed to be trapped by the laser beam, since gradient forces overcome radiation pressure and radiometric forces \cite{Svoboda1,Svoboda2,Rocha,Campos}. OT have applications in areas such as biochemistry, biophysics, microfluidics, colloidal sciences and others, allowing mechanical studies of small soft matter systems \cite{Ashkin444,Grier,Svoboda1,Neuman,beugnon2007two,Moffitt,Neuman1,kress2009cell,wu2011optoelectronic%
,Fazal,capitanio2012ultrafast,phillips2014shape}. From an electric conducting point of view, semiconductor materials interpolate between dielectrics and metals, being the main ingredients for microeletronics technology. Consider shedding light onto a semiconductor microparticle. What dynamics could we expect for it? Would it be trapped, like a dielectric, or drifted away, as occurs to metallic beads? To address these questions we have conducted an experimental study upon the optical trapping and manipulation of Germanium (Ge) microspheres (diameter around a few micrometers) under the action of a highly focused Gaussian laser beam OT. Surprisingly, Ge beads are observed to oscillate in a plane perpendicular to the optical axis with relatively well-defined amplitude and frequency controllable by the laser power. Fig. \ref{Fig1}(a) shows our experimental setup, in which the particles are located around $z \sim -5\,\mu$m below the focal plane. The position of a $3\,\mu$m diameter Ge particle relative to the optical axis is shown in the inset, illustrating their typical oscillatory dynamics. In addition, we show that laser polarization can be used to easily guide the oscillations to a preferential direction, a very useful feature for practical applications. Such an oscillatory motion has the potential of extending usual OT setup capabilities for investigating dynamical properties of macromolecules and small systems, like DNA molecules and biological membranes.

Over 100 years after its discovery, Ge continues to attract tremendous attention in different areas of condensed matter physics and materials science \cite{Dyakonov,Pezzoli,Dushenko,Bottegoni,Sze,Pillarisetty}. Due to its suppression of spin relaxation, the group IV semiconductors, such as Ge and Si, has aroused interest in timely research branches, like spintronics. Nowadays, Ge has application in optical fibers, polymerization catalysts, and Si-Ge alloys in microchip manufacturing, with feature sizes on the chips reaching 7 nm ($<$ 60 Ge atoms) \cite{Pillarisetty,Burdette}. However, to our best knowledge, the trapping of Ge microparticles in OT has not been investigated so far. Our Ge-microparticles have been prepared using laser ablation technique in liquid solution (LATLS). Their high quality has been confirmed by Scanning Electron Microscope (SEM) and Raman scattering [see Fig. \ref{Fig2}(b-c)]. Additional characterizations of energy dispersive X-ray (EDX) analysis have also demonstrated that the synthesized particles have no impurities in their composition.

\begin{figure*}
\includegraphics[width=\linewidth]{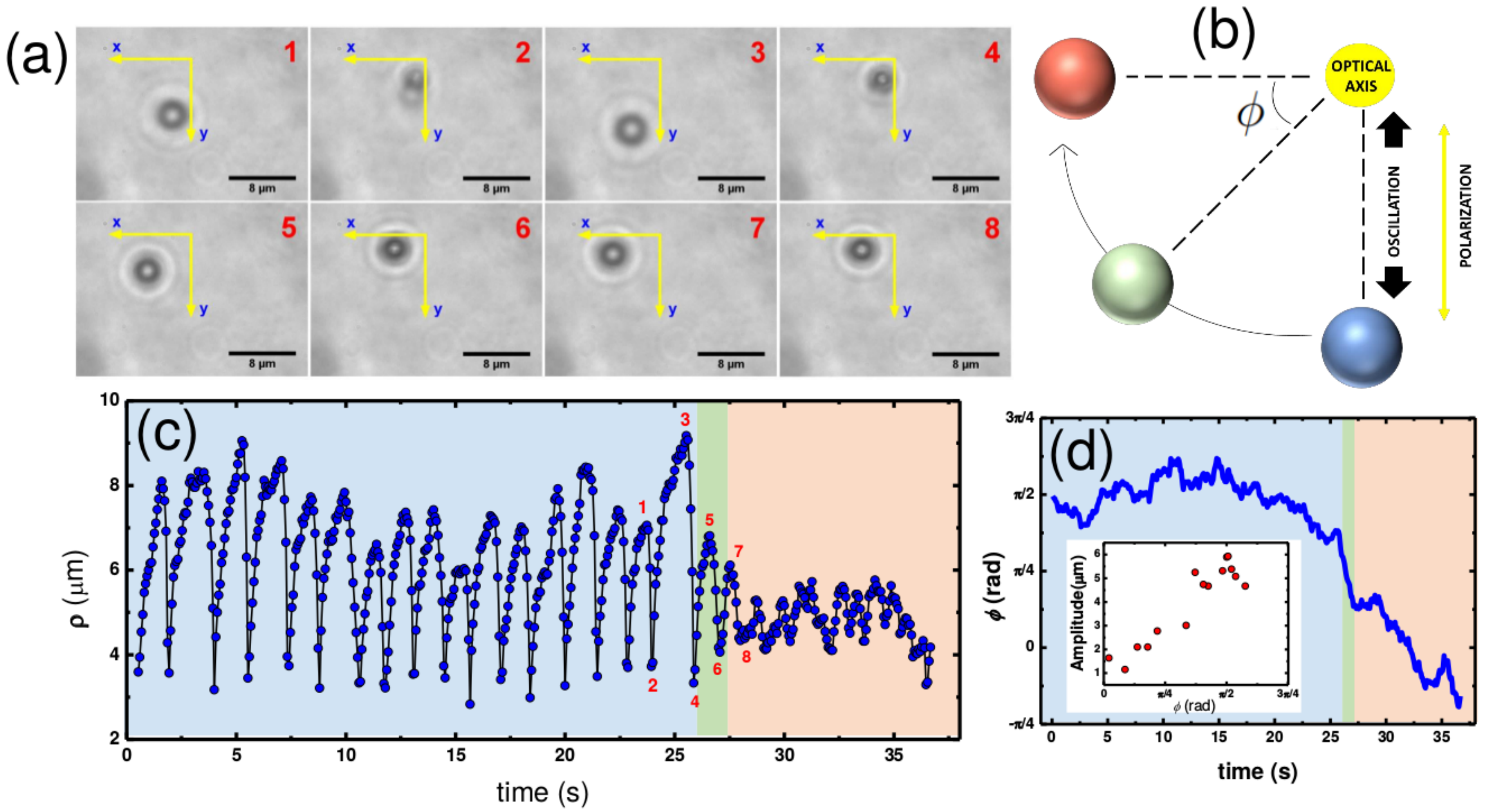}
\caption{(Color online) Ge microparticle of $3.1\,\mu$m diameter under a linearly polarized  Gaussian laser beam OT. (a) Successive video frames showing the microsphere oscillating toward the optical axis (origin of the coordinate frame). (b) Direction of light polarization, along with the definition of the azimuthal angle, $\phi$. (c) Radial displacement of the particle, $\rho$, as function of time. The blue (red) background indicates oscillations along the $y$-axis ($x$-axis), while the green background points out the transient from one axis to the other. (d) How the azimuthal angle, $\phi$, varies in time. The inset shows the amplitude of the oscillations vs. $\phi$.}
\label{Fig3}
\end{figure*}

For the optical experiment, we have selected those particles with well-defined spherical shape, and set the laser power to 37 mW, measured at the objective entrance. Fig. \ref{Fig3} shows a number of features of the oscillatory motion of a $3.1\,\mu$m diameter Ge-microsphere under a linearly polarized and highly focused Gaussian laser beam. Let us define a cylindrical coordinate system whose origin is at the laser beam focus, while $y$-axis takes along the direction of polarization [see successive video frames in Fig. \ref{Fig3}(a)]; let $z$-axis be along the optical axis and parallel to laser propagation. The position of the particle is represented by $\vec{r}=(\rho, \phi, z)$, where $\rho$ is the radial particle displacement relative to the optical axis, while $\phi$ is the azimuthal angle, measured relative to the $x$-axis [see Fig. \ref{Fig3}(a-b)]. How the particle radial position, $\rho(x,y)=\sqrt{x^2+y^2}$, varies over time is plotted in  Fig. \ref{Fig3}(c). Notice that the particle is initially in a position parallel to the light polarization, say, along $y$-axis ($\phi=\pi/2$), where it begins to oscillate toward the optical axis. However, the oscillation direction changes with time, eventually reaching a direction perpendicular to the polarization axis, as shown in Fig. \ref{Fig3}(d). Once there, the oscillatory direction is relatively stable around $x$-axis ($\phi=0,\pi$). The amplitude of oscillation appears to increase linearly with the azimuthal angle, $\phi$, in the range of $[0,\pi/2]$ [see inset of Fig. \ref{Fig3}(d)]. In addition, the diffraction pattern of the microparticle changes periodically in time [see Video.avi (Supplemental Material) and Fig. \ref{Fig3}(a)], indicating that it is also subject to oscillations along the direction of the laser propagation, a behavior predicted in Ref. \cite{Campos2}.

As briefly mentioned before, the oscillatory dynamics is driven by the competition between gradient and radiometric forces. The so-called gradient force, $\vec{F}_G$, is in order due to refraction of light at the particle surface, and acts as an attractive force that pulls the bead toward the optical axis. When generated by a linearly polarized Gaussian laser beam, the components of the gradient force in our effective model acquires additional factors with simple dependence on $\phi$. Within the Mie scattering formalism, these factors arises from the asymmetry imposed by the linear polarization \cite{Dutra}. See methods for an analytical expression for $\vec{F}_G$ and Supplemental Material \cite{SupMat} for a detailed construction.

On the other hand, absorption of light by the microsphere leads to the asymmetric heating of the particle and its surrounding medium, giving rise to the radiometric force, $\vec{F}_R$. The radiometric force pushes the particle from the focal region, and can be approximated as proportional to the laser intensity \cite{Campos,Ke}. See the expression for $\vec{F}_R$ at methods section and further discussion in the Supplemental Material \cite{SupMat}.

The resultant force exerted on a microsphere is given by  $\vec{F}=\vec{F}_G+\vec{F}_R$. While the radiometric force tends to push the particle away, gradient force works as a restoring action towards the laser beam focus. Depending on the optical properties of the material from which the particle is made, three main scenarios emerge: 1) Radiometric force dominates: as occurs to metallic beads, large absorption of the incident light results in inhomogeneous heating of the medium and the particles are drifted away; 2) Suitable conditions favoring optical trapping are accomplished: for example, a dielectric microsphere in which absorption and reflection of light are negligible and gradient force dominates; 3) Special situation takes place whenever both of such forces have comparable magnitudes: the competition between them may yield oscillatory motion, as we have observed to occur with Ge-microspheres. Actually, similar oscillations also happen with topological insulator Bi$_2$Te$_3$ and Bi$_2$Se$_3$ beads \cite{Campos}. Once such a kind of compounds share some electrical and optical properties with semiconductors, we may wonder whether optically induced oscillatory motion is the typical dynamics of semiconductor beads under the action of focused laser light, as commonly used in OT setup.

\begin{figure*}
\center
\includegraphics[width=.85\textwidth]{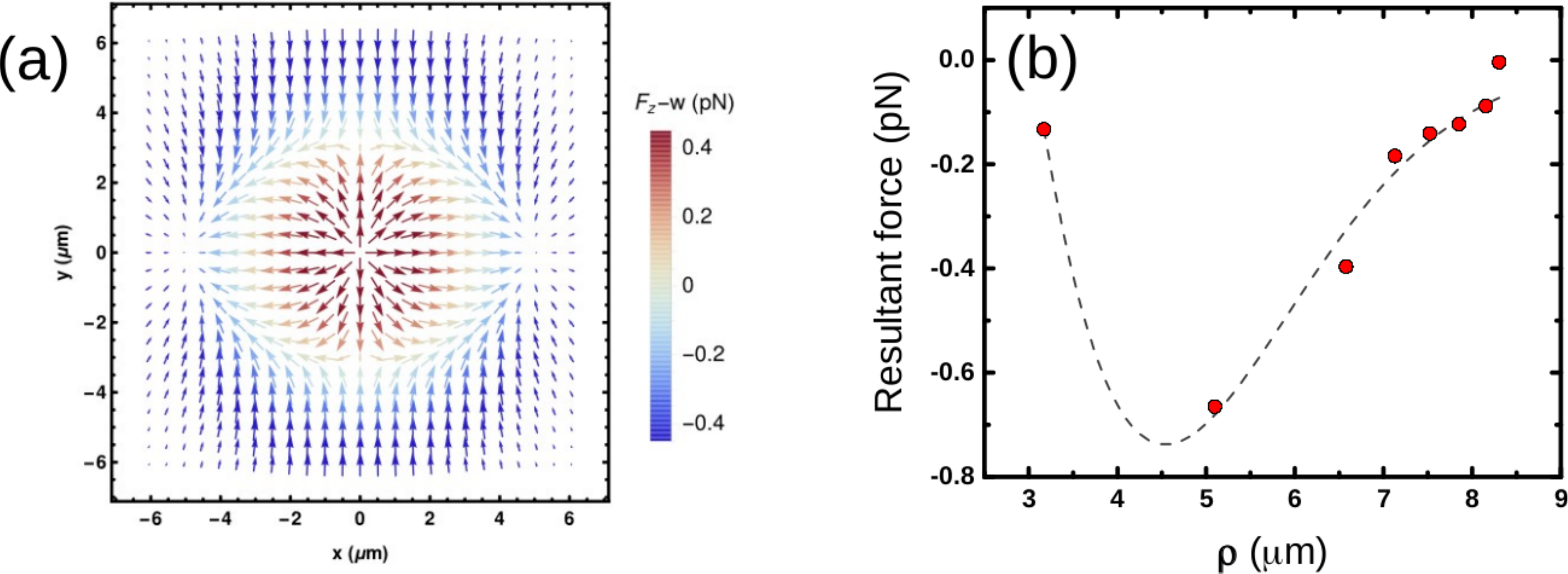}
\caption{(Color online) (a) Typical vector plot of the resultant force, $\vec{F}=\vec{F}_R+\vec{F}_G$,  for the approximation regime. The in-plane vectors where plotted using the $x$ and $y$ components of the resultant force, while the color bar represents the total force along $z$-direction, $F_z-{\rm W}$, where $\vec{W}=-(4/3)\pi a^3(\rho_{Ge}-\rho_m)g  \hat{z}$ is the apparent weight of the particle \cite{Campos2}. $F_z-{\rm W}>0$ for red vectors, while blue ones stands for $F_z-{\rm W}<0$. $a\approx1.55\,\mu{\rm m}$ is the sphere radius, $g=9.8 {\rm m/s^2}$ is the acceleration of gravity, $\rho_{Ge}=5323\,{\rm kg/m^3}$ and $\rho_m=997\,{\rm kg/m^3}$ are the Ge and medium (deionized water) densities, respectively. We have taken the following numeric values for the parameters: $w(z)=5.2\,\mu$m, $\mathcal{F}_{g\rho}=3.5\,$pN, $\mathcal{F}_{g\phi}=0.7\,$pN, $\eta_\rho=0.22$, $\mathcal{F}_{r\rho}=7.0\,$pN, $\mathcal{F}_{rz}=1.3\,$pN and $\eta_z=0.1$. (b) Typical radial component of the resultant force as function of the distance from the optical focus, $\rho$. Red dots are experimental data extracted from the third oscillatory cycle in Fig. \ref{Fig3}-(c), which takes place mostly along the polarization direction ($y$-axis, in which the azimuthal component, $F_{\phi}$, is negligible). The dotted line is the fitting of the radial component of the resultant force with data from the approximation regime. The fitting parameters along with their standard errors read as: $\mathcal{F}_{r\rho}=(7.0\pm1.4)\,$pN, $\mathcal{F}_{rg}=(3.59\pm0.56)\,$pN and $w(z)=(5.20\pm0.18)\,\mu$m.}
\label{vectorplot}
\end{figure*}
% % % % % %

Let us discuss the physical effects underlying the oscillations by considering two distinct regimes: \emph{Approximation regime} refers to the time interval in which the particle is moving toward the optical axis [e.g. from 1 to 2 in Fig. \ref{Fig3}(c)]. The particle is initially at maximum distance from the optical axis, where gradient force dominates. Then, it moves toward the region of higher intensity and the temperature increases very fast around a particular distance to the optical axis. As a consequence, radiometric force overcomes the gradient and the transition to the removal regime takes place very abruptly, resulting in sharp peaks in the particle position evolution at the points of minimum radial distance ($\rho\sim 3\,\mu$m in Fig. \ref{Fig3}(c)). \emph{Removal regime} refers to the time interval in which the particle is moving away from the optical axis [e.g. from 2 to 3 in Fig. \ref{Fig3}(c)]. The particle is initially at minimum radial distance and repulsive radiometric force dominates. The particle is pushed away from the optical axis in such a way that the radiometric force overcomes the gradient along all the removal regime (for more details about the resultant force in the removal regime, see Fig. S2 and related discussion in the Supplemental Material \cite{SupMat}). The particle cools down until it reaches thermal equilibrium at the point of maximum distance. Thus, smooth transition to the approximation regime takes place, restarting the cycle with the gradient force dominating the dynamics.

A precise description beyond our effective model is much more involved, once the physical parameters comprised in the resultant force, Eqs. (\ref{gradient-force2}-\ref{radiometric-force}), are expected to bear complicated dependence on the microsphere and medium optical properties at each time instant. In order to simplify our analysis, we shall consider constant effective values for the model parameters ($\mathcal{F}_{G\rho}$, $\mathcal{F}_{R\rho}$, etc) for the approximation and removal regimes separately. Fig. \ref{vectorplot}(a) shows the vector plot of the resultant force, $\vec{F}$, for the approximation regime, at a cross section of height $z=-5.0\,\mu$m. Note that there are two ``stable'' points along the direction perpendicular to the polarization axis, so that the particle will be eventually attracted to one of them. However, the time required for the particle to reach out a stable point is longer than the oscillation period, so that a particle initially placed at the $y$-axis performs a few oscillations before displacing to $x$-axis. Along this direction the amplitude of the oscillations decreases whenever compared with $y$-axis, by virtue of polarization effect.

Fig. \ref{vectorplot}(b) shows the radial component of the resultant force, $F_{\rho}$, as function of the radial position, $\rho$, for the approximation regime of the third oscillatory cycle in Fig. \ref{Fig3}(c). The dashed line represents a theoretical fitting of $F_\rho$ with experimental data. Among the fitting parameters, we have $w(z\approx-5\,\mu\textrm{m})=(5.20\pm0.18)\,\mu$m. From $w(z)$, the experimental parameter $w_{0-exp}=(0.327\pm 0.011)\,\mu$m is readily obtained, whose value is in good agreement with its theoretical counterpart $w_0=0.36\,\mu$m. [Taking the average over many cycles, we have obtained $w_{0-exp}=(0.31\pm 0.02)\,\mu$m].

\begin{figure}
\includegraphics[width=\linewidth]{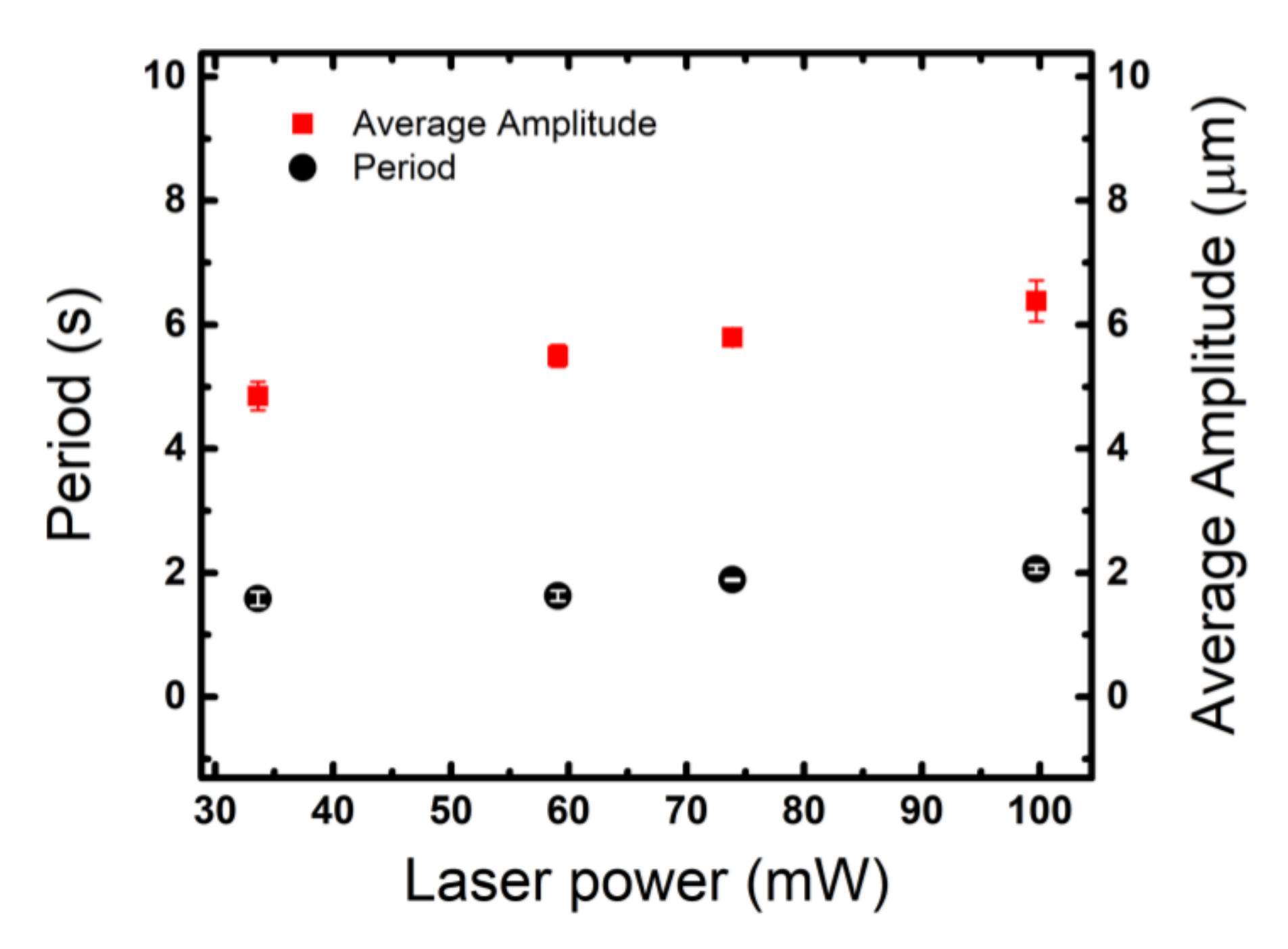}
\caption{(Color online). Period (black dots) and amplitude (red squares) of oscillation as laser power is varied.  Vertical bars accounts for standard errors over the averages.}
\label{per-amp-vs-power}
\end{figure}

Fig. \ref{per-amp-vs-power} shows the dependence on the laser power of the period and average amplitude of oscillations. Note that the amplitude increases linearly with the laser power and the period also exhibits a slight increase in such range. Therefore, the laser power can be used to control these dynamic quantities, which is essential when considering practical applications.

\vskip 1cm
Finally, we believe that the opto-thermal oscillations of the Ge microparticles constitute a very rich scenario for the experimental realization of a single particle thermal machine (microscopic engine) \cite{MT1,MT2,martinez2017colloidal}. It has been recently reported the realization of a Brownian Carnot engine in which the working substance is a single polystyrene colloidal particle trapped in an OT \cite{MT3}. However, important features make thermal machines with Ge particles a promising alternative. 1) Robust dynamics: it depends only on the optical properties of the particles and the medium. No external perturbation is necessary in order to achieve a thermodynamic cycle. 2) Injection of heat takes place via absorption of light, so the laser beam itself plays the role of the hot bath. A significant amount of energy that is usually wasted in conventional traps is converted into work in such system. 3) The large amplitude of oscillations is exceptionally appealing when it comes to practical applications of a thermal machine, such as exerting dynamic forces on polymers and other microscopic objects.

In summary, Ge microparticles suspended in water are observed to oscillate whenever subject to a highly focused laser beam, with remarkable similarity to what occur to Bi$_2$Te$_3$ and Bi$_2$Se$_3$ topological insulator beads \cite{Campos}. Once such a kind of compounds share some electrical and optical properties with semiconductors, we may wonder whether optically induced oscillatory motion is the typical dynamics of semiconductor beads under the action of focused laser light. More specifically, Ge-particles tend to oscillate in a direction perpendicular to the linear polarization of the laser beam. A possible explanation for such a dependence on the laser polarization is based on the theory of charge carriers in crystal lattice materials: Ge has a considerable density of free carriers inside its bulk, rendering a strong response to the light polarization. In contrast, topological insulators compounds have most of their free electrons lying on the surface, due to the energy band gap in the bulk of the material. This would justify why the topological insulator beads are practically insensitive to light polarization, as reported in Ref. \cite{Campos}. Our present findings suggest that semiconductor-type beads tend to oscillate under the action of highly focused laser beams, as those used in OT setups. In fact, we also verified the signature of this behavior in CdTe particles, which indicates that such a novel phenomenon may be typical of conventional semiconductor materials. Therefore, these materials could also find application for dynamical measurements of mechanical properties of small systems, including macromolecules such as DNA, membranes and so forth.

\section*{Methods}

In this section we provide a brief description of the methods and effective theoretical model adopted in this work. For more details, see the Supplemental Material \cite{SupMat}.\\
\textbf{Synthesis and characterization of the Ge microparticles}. The particles of Ge were synthesized by pulsed laser ablation technique in liquid solution. We have used bulk Ge-crystals as a target material immersed in distilled water. The laser beam was focused onto the Ge surface and the target was moved perpendicularly to the laser beam to irradiate fresh surfaces during the whole process. The high quality of the Ge particles has been confirmed by Scanning Electron Microscope and Raman scattering. Additional characterizations of energy dispersive X-ray analysis have also demonstrated the absence of impurities in the particles composition.\\
\textbf{Optical Setup}. The OT consist of a 1064 nm ytterbium-doped fiber laser (IPG Photonics) operating in the TEM$_{00}$ mode, mounted on a Nikon Ti-S inverted microscope with a 100$\times$ NA 1.4 objective. We have set the laser power at 37 mW, measured at the objective entrance (exception is made for those results depicted in Fig. 5 and related text, where frequency and amplitude are studied under power variation). The sample chamber consists of an o-ring glued in a microscope coverslip.\\
\textbf{Data analysis}. The individual motion of each Ge microparticle for various oscillation cycles was recorded using videomicroscopy. The videos were analyzed using the ImageJ software, which allows us to obtain the coordinates of the particle centroid as a function of time. The velocity and acceleration necessary to determine the Stokes and resultant forces are calculated by taking the first and second order time derivative of the particle position, respectively.\\
\textbf{Effective model for gradient and radiometric forces}. Here we present the equations for the gradient and radiometric forces exerted on a Ge particle under a linearly polarized Gaussian laser beam optical tweezers. For an extended discussion upon the derivation of the effective theoretical model, see the corresponding section in the Supplemental Material \cite{SupMat}.

The gradient force components, $\vec{F}_G= F_{G\rho}\hat{\rho}+ F_{Gz}\hat{z}+ F_{G\phi}\hat{\phi}$, reads:
\begin{eqnarray} \label{gradient-force2}
& F_{G\rho}=& -\dfrac{2\rho\mathcal{F}_{G\rho}\exp(1/2)}{w(z)(1+\eta_\rho)}\exp\left(\frac{-2\rho^2}{w(z)^2}\right)\left(1-\eta_\rho\cos 2\phi\right), \hskip .5cm\\ \label{gradient-force3}
& F_{Gz}=& \frac{2\mathcal{F}_{Gz}\rho^2}{w(z)^2(1+\eta_z)}\exp\left(\frac{-2\rho^2}{w(z)^2}\right)\left(1-\eta_z\cos 2\phi\right)\,,\\ \label{gradient-force4}
& F_{G\phi}=&-\frac{2\rho\mathcal{F}_{G\phi}\exp(1/2)}{w(z)}\exp\left(\frac{-2\rho^2}{w(z)^2}\right)\sin 2\phi \,.
\end{eqnarray}%
The factor $\exp\left(\frac{-2\rho^2}{w(z)^2}\right)$ is the normalized intensity of a Gaussian laser beam, where $w(z)=w_0\sqrt{1+(z/z_R)^2}$ is the beam waist at height $z$ relative to the focal plane, and $\quad z_R\equiv\pi w_0^2/\lambda$. $w_0$ is the beam waist at the focal plane and $\lambda$ is the light wavelength. The parameters $\mathcal{F}_{G\rho}, \mathcal{F}_{Gz}$, and $\mathcal{F}_{G\phi}$ are the maximum magnitude of each component, whose highest values occur at $(\rho,\phi) =(w(z)/2 , \pm\pi/2),\,(w(z)/\sqrt{2}, \pm\pi/2)$, and $(w(z)/2, \pm \pi/4, \pm 3\pi/4, \pm 5\pi/4,\ldots)$. In turn, $\eta_\rho$ and $\eta_z$ account for the asymmetries brought about by linear polarization direction.

The simplest expression for describing the radiometric force is given by (for discussion, see Supplemental Material \cite{SupMat}) \cite{Ke,Campos}:

\begin{equation}
\vec{F}_R={F}_{R\rho} \hat{\rho} + {F}_{Rz}\hat{z} = (\mathcal{F}_{R\rho} \hat{\rho}+\mathcal{F}_{Rz}\hat{z})\exp\left(\frac{-2\rho^2}{w(z)^2}\right). \label{radiometric-force}
\end{equation}
$\mathcal{F}_{R\rho}$ and $\mathcal{F}_{Rz}$ are the maximum magnitudes of the radial and $z$ components of the radiometric force, respectively, corresponding to their values at the optical axis. Eq. (\ref{radiometric-force}) can also account for the radiation pressure force, which is important in some cases, depending on the reflectance of the material.

\section*{Acknowledgements}
The authors are grateful to Marcelo L. Martins for insightful discussions regarding single particle thermal machines and Libor \v{S}mejkal for reviewing the manuscript. This study was financed in part by the Coordena\c c\~ao de Aperfei\c coamento de Pessoal de N\' ivel Superior - Brazil (CAPES) - Finance Code 001. The authors also thank CNPq, FINEP and FAPEMIG (Brazilian agencies) for financial support.

\bibliography{GeArticle.bib}

\end{document}